\title{Quasi-steady description of modulation effects in wall turbulence}
\author
{Sergei I.~Chernyshenko$^1$\thanks{Email address for correspondence: s.chernyshenko@imperial.ac.uk}, Ivan Marusic$^2$,\\
 and Romain Mathis$^2$\\
\small $^1$Department of Aeronautics, Imperial College, London, SW7 2AZ, UK\\
\small $^2$Department of Mechanical Engineering, University of Melbourne,\\
\small Victoria 3010, Australia
}

\documentclass[12pt]{article}

\usepackage{graphicx, psfrag}
\usepackage{natbib}
\usepackage{
amssymb, amsmath, wasysym}
\usepackage{hyperref}

\newcommand{\etal}{{\itshape et al.}}
\newcommand\Rey{{\mathrm Re}}

\begin{document}
\maketitle
\begin{abstract}
A theoretical description of the phenomenon of modulation
of near-wall turbulence by large scale structures is investigated.
The description given is simple in that the effect of large-scale structures is limited to a
quasi-steady response of the near-wall turbulence to slow large-scale
fluctuations of the skin friction. The most natural and compact form
of expressing this mechanism is given by the usual
Reynolds-number-independent representation of the total skin friction
and velocity, scaled in wall variables, where the mean quantities are
replaced by large-scale low-pass-filtered fluctuating
components. 
The theory is rewritten in terms of fluctuations via a universal mean velocity
 and random mean square fluctuation velocity profiles of the small-scales and then
 linearised assuming that the large-scale fluctuations are small as compared to the mean components.
 This allows us to express the superposition and modulation
coefficients of the empirical predictive models of the skin friction
and streamwise fluctuating velocity given respectively by
Marusic~\etal~($13^{th}$ {\it Eur. Turb. Conf.}, 2011) and
  Mathis~\etal~({\it J. Fluid Mech.} 2011, vol. 681, pp. 537--566). 
It is found that the theoretical quantities agree well with
experimentally determined coefficients.
\end{abstract}

\section{Introduction}

Direct numerical simulations of turbulent skin friction drag reduction
are necessarily done at moderate Reynolds numbers ($\Rey$). 
Until recently, the hope that theories justified by comparisons with such
calculations can be extrapolated to much larger $\Rey$  typical for
flight regime, was based on the idea of the universality of near-wall
turbulence, where the skin friction is primarily generated. 
This hope suffered a serious blow with the discovery of the emergence,
at sufficiently large $\Rey,$ of a secondary peak in the pre-multiplied
energy spectra map of the streamwise velocity component, namely the
``outer-peak''~\cite[see for example][]{HutchinsMarusic2007a}. 
Empirical fits based on the latest data
(\citealt{Alfredsson2011ETC13}) suggest that at high $\Rey$  more
turbulence kinetic energy is contained in large-scale structures than 
in the near-wall region where the inner-peak is located. 
In the past decade or so, the large-scale structures and their
interaction with near-wall turbulence have attracted much attention 
(\citealt{AdrianMeinhartTomkins2000};
\citealt{delAlamoJimenezPF2003};
\citealt{BharathJFM2003};
\citealt{delAlamoJimenezEtAlJFM2004};
\citealt{GanapathisubramaniEtAlJFM2005};
\citealt{HoyasJimenezPF2006};
\citealt{HambletonHutchinsMarusicJFM2006};
\citealt{McKeonMorrisonPTRS2007};
\citealt{MorrisonPTRS2007}, and even this list is far from complete).
In particular, it has been shown that the large-scale structures
associated with the log-layer  significantly affect the near-wall
turbulence. 
\cite{AbeKawamuraChoiASMEFluidEng2004} and \cite{HutchinsMarusic2007a}
have observed that these large-scale motions impose a substantial
``footprint'' all the way down to the wall, and this has been confirmed in recent direct numerical simulations (DNS) of evolving boundary layers (\"Orl\"u \& Schlatter 2011). \nocite{OrluSchlatter2011}
This footprint is seen as 
a long-wavelength component superimposed onto the small-scale
near-wall fluctuations. Further, \cite{HutchinsMarusic2007b} and
\cite{MathisHutchinsMarusicJFM2009}  demonstrated that the
large-scale motions are not merely superimposed near the wall, but
also amplitude modulate the small-scale motions, and developed a tool to quantify the degree of 
modulation.
Based on the above observations, Marusic and colleagues developed
a simple quantitative model of the effect of large-scale structures
onto the near-wall turbulence (Marusic \etal~2010; Mathis \etal~2011), and from here on we refer to these papers as MHM.
In this model the effect of the large-scale structures is expressed as a
combination of an amplitude modulation and superposition of the large-scales 
onto the near-wall small-scales. 
The amplitude modulation factor and the superimposed additive term
were approximated as linear functions of the low-pass-filtered
velocity in the logarithmic layer. 
The model was shown to reconstruct a statistical representation of the
near-wall fluctuating streamwise velocity component, as well as the
fluctuating skin-friction, with good accuracy over a large range of
$\Rey.$  

\nocite{marusic10_SCI,MathisJFM2011,MarusicETC}

In the present paper we demonstrate that the modulation effect as observed in an experiment can be derived from a simple hypothesis.

\section{Skin friction modulation}\label{sec:SkinFriction} 

The classical view on the universality of near-wall turbulence consists in the statement that all the flow variables, if expressed in wall units, are independent of the Reynolds number. For skin friction $\tau$ this means that
\begin{equation}\label{eqn:UniversalityOfTau}
\tau = \bar\tau \tau^*\left(\frac{t\bar u_\tau^2}{\nu},\frac{x\bar u_\tau}{\nu},\frac{y\bar u_\tau}{\nu}\right),
\end{equation}
where $\tau^*(t^+,x^+,y^+)$ is a universal function of its arguments in the sense that all its statistics are independent of $\Rey,$ $\bar \tau$ is the mean skin friction,  $\bar u_\tau=\sqrt{\bar\tau/\rho},$ $\nu$ is the kinematic viscosity, $\rho$ is the density, and $t,$ $x,$ and $y$ are time and coordinates in the plane of the wall. Note that all the quantities here are the total values: in this paper fluctuations will always be explicitly marked with prime or tilde. Note that the time average of $\tau^*$ equals one: $\overline{\tau^*}=1.$

The experimentally observed effect of large-scale structures on the near wall turbulence is in contradiction with (\ref{eqn:UniversalityOfTau}). We propose therefore to replace (\ref{eqn:UniversalityOfTau}) with a formula recognising the dependence of the skin friction on the large scale effects. Our proposal might be reminiscent of the transition from the famous Kolmogorov-41 theory to Kolmogorov-62 theory in the isotropic turbulence, but we will not further comment on this similarity.

The proposed hypothesis consists in the statement that the effect of large scale structures on the near-wall turbulence is limited to replacing $\bar \tau$ (and, correspondingly, $\bar u_\tau$) in (\ref{eqn:UniversalityOfTau}) with slowly-varying $\tau_L(t)$ (and $u_{\tau_L}(t)$): 
\begin{equation}\label{eqn:TauModulated}
\tau = \tau_L(t) \tau^*\left(\frac{tu_{\tau_L}^2(t)}{\nu},\frac{xu_{\tau_L}(t)}{\nu},\frac{yu_{\tau_L}(t)}{\nu}\right).
\end{equation} 
The slowly-varying component of the wall skin friction can be extracted from the full skin friction by a suitable filtering procedure.
A quasi-steady effect described by (\ref{eqn:TauModulated}) can naturally be called a modulation of near-wall turbulence by large-scale structures. We will now try to verify the hypothesis (\ref{eqn:TauModulated}) by comparing it with the available experimental data.  

Note that in Marusic \etal~(2011) the modulation was understood somewhat
differently, and it was described in terms of fluctuations.  
To compare, we rewrite (\ref{eqn:TauModulated}) in terms of
fluctuations, with $\tau^*=1+{\tau'}^*$ (for brevity we will omit
arguments of $\tau^*$). 
Assuming that $\tau_L$ and $\tau^*$ are not correlated, we also have
$\overline{\tau}=\overline{\tau_L}$. Then
$\tau_L=\overline\tau+\tau_L'$ and $\tau=\overline\tau+\tau'$.
Substituting these formulae into (\ref{eqn:TauModulated}) and
rearranging gives 
\begin{equation}\label{eqn:TauPrimeModulated}
{\tau'}^+=\frac{\tau'}{\overline\tau} = \left(1+\frac{\tau_L'}{\overline{\tau}}\right){\tau'}^*+\frac{\tau_L'}{\overline{\tau}}.
\end{equation} 

Compare this with equation (1) in Marusic \etal~(2011), which has the form
\begin{equation}\label{eqn:MarusicETCEqOne}
{\tau'_p}^+= 
{\left(1+\beta {u'}_{OL}^+\right)\tau'}^*
+\alpha {u'}_{OL}^+,
\end{equation} 
where ${u'}_{OL}^+$ is the fluctuating large-scale signal from the log
region, and we added primes to fluctuation quantities, to bring it
closer to our notation (subscript ``$p$'' describing a predicted
quantity). 

One can see that if our explanation of the modulation is true, then 
$$\frac{\tau_L'}{\overline{\tau}}=\alpha {u'}_{OL}^+$$ and $$\frac{\tau_L'}{\overline{\tau}}=\beta {u'}_{OL}^+,$$ that is $\alpha$ and $\beta$ should be equal. In
Marusic \etal~(2011) these quantities were found independently from experiment using very different procedures, and it was obtained that $\alpha=0.0898$ and $\beta=0.0867.$ The small difference between these values gives the first confirmation of our interpretation of the modulation effect.

\section{Velocity modulation} 

The classical view on the universality of velocity distribution gives:
$$
u=\sqrt{\frac{\bar\tau}{\rho}}u^*=\bar u_\tau u^*\left(\frac{t\bar u_\tau^2}\nu,\frac{x\bar u_\tau}\nu,\frac{y \bar u_\tau}\nu,\frac{z\bar u_\tau}\nu\right)
$$
The idea of modulation can now be expressed again as a statement that the effect of large scale structures on the near-wall velocity distribution is limited to introducing into the above formula a dependence of $\bar \tau$ or, more conveniently in this case, the corresponding dependence of $\bar u_\tau$ on time, with the related assumption that the variation of these quantities is much slower than the variation of $u^*:$

\begin{equation}\label{UniversalTotalVelocity}
u=u_{\tau_L}(t) u^*\left(\frac{tu_{\tau_L}^2(t)}\nu,\frac{xu_{\tau_L}(t)}\nu,\frac{yu_{\tau_L}(t)}\nu,\frac{zu_{\tau_L}(t)}\nu\right)
\end{equation}

 Accordingly, this means that the large scale structure effect amounts to the amplitude modulation of the universal total (not fluctuating) velocity $u^*$ via the factor $u_\tau=u_{\tau_L}(t),$ a frequency modulation due to what in fact is the time-dependence of the wall units in the expression $tu_{\tau_L}^2(t)/\nu,$ and a scale modulation due to the time-dependence of the wall units in $xu_{\tau_L}(t)/\nu,$ $yu_{\tau_L}(t)/\nu,$ and $zu_{\tau_L}(t)/\nu.$ In treating the skin friction in Section~\ref{sec:SkinFriction}, as far as the random mean square of the skin friction fluctuation was concerned, the frequency and scale modulation could be ignored because the skin friction random mean square is actually independent of time and spatial coordinates. In the present case, this is true only for the frequency and scale modulation in the wall-parallel directions. When, as a result of the variation of $u_{\tau_L}$ the wall-normal non-dimensional distance $zu_{\tau_L}(t)/\nu$ varies, so does the random mean square. For this reason in what follows we will frequently omit the   $tu_{\tau_L}^2(t)/\nu,$ $xu_{\tau_L}(t)/\nu,$ and $yu_{\tau_L}(t)/\nu$ arguments of $u^*,$ but we will often keep $zu_{\tau_L}(t)/\nu.$ 

It is also convenient to introduce new variables reflecting the frequency and scale modulation, namely
$$
T^+=\frac{tu_{\tau_L}^2(t)}{\nu},\qquad Z^+=\frac{zu_{\tau_L}(t)}{\nu}.
$$
Note the difference between $T^+$ and $Z^+$ and the variables in wall units, defined similarly but with the time-averaged value of the friction velocity $\bar u_\tau=\overline{u_{\tau_L}(t)}:$
$$t^+=\frac{t\bar u_\tau^2}{\nu},\qquad z^+=\frac{z\bar u_\tau}{\nu}.$$

Our goal is to rewrite (\ref{UniversalTotalVelocity}) in fluctuations and to compare it with the formula given in MHM.

Let us introduce the universal mean profile as
\begin{equation}\label{FastMeanU}
U^*(z^+)=\lim_{T\to\infty}\frac1{2T}\int_{-T}^{+T}u^*(t^+,z^+)\,dt^+
\end{equation}
and the universal velocity fluctuation function 
\begin{equation}\label{eqn:UniversalFluctuation}
\tilde u^*(t^+,z^+)=u^*(t^+,z^+)-U^*(z^+).
\end{equation}
Note that both the average and the fluctuation are introduced here with $z^+$ held constant as time varies. Where the velocity at a fixed distance $z$ to the wall will be considered, the argument of $u^*$ will be $Z^+$, which is varying in time, and not $z^+.$ This is why a tilde and not a prime is used to mark a fluctuation in (\ref{eqn:UniversalFluctuation}).

Note now the due to the separation of time scales of near-wall turbulence and large scale structures 
$$
\overline{u^*(T^+,Z^+)}=\overline{U^*(Z^+)}.
$$ 
The assumption of the separation of time scales can be formalised and then the above statement can be proved formally, but in this paper we will focus at the level of relating to the physical understanding. Introducing the fluctuation of the friction velocity $u_{\tau_L}' = u_{\tau_L}(t)-\bar u_\tau, $ one can write
$$
u=(\bar u_\tau+u_{\tau_L}'(t))\left(U^*\left(Z^+\right)+\tilde u^*\left(T^+,Z^+\right)\right)
$$
Due to the separation of scales
$$
\overline{u_{\tau_L}'(t)\tilde u^*\left(T^+,Z^+\right)}=0.
$$
Hence the velocity average is
$$
\overline u=\bar u_\tau \overline{U^*\left(Z^+\right)}+\overline{u_{\tau_L}' U^*\left(Z^+\right)}.
$$ 
and its fluctuation, expressed in wall units, is
\begin{equation}\label{eqn:FullUprimeModulationA}
{u'}^+=\frac{u-\bar u}{\bar u_\tau}= \tilde u^*\left(T^+,Z^+\right)\left(1+\frac{u_{\tau_L}'(t)}{\bar u_\tau}\right)+U^*\left(Z^+\right)\frac{u_{\tau_L}'(t)}{\bar u_\tau}+U^*\left(Z^+\right)-\frac{\bar u}{\bar u_\tau}
\end{equation}

The corresponding formula in MHM has the form
\begin{equation}\label{eqn:MathisEqTwoPointOne}
{u'_p}^+=u^*_{\mathrm{MHM}}\left\{1+\beta_u {u'}_{OL}^+\right\}+\alpha_u {u'}_{OL}^+,
\end{equation} 
in which we include primes where appropriate, and have added the subscript $u$ to $\alpha$ and $\beta$ to distinguish them from $\alpha$ and $\beta$ of Section~\ref{sec:SkinFriction}.

Both these formulae represent the velocity fluctuation as a fast varying term multiplied by a slowly varying (depending only on large scale structure) amplitude modulation term plus a slowly varying superposition term. We might now try to associate $u_{\tau_L}'(t)/ \bar u_\tau$ with ${u'}_{OL}^+$ similar to Section~\ref{sec:SkinFriction}. However, there are two difficulties here. First, in MHM $\beta_u$ was chosen in such a way that $u^*_{\mathrm{MHM}}$ is not amplitude-modulated by the large scale structures. More precisely, $\beta_u$ is chosen in such a way as to minimize the effect of large scale structures on the random mean square of the fast-varying term. This can be reproduced in our formula by introducing the mean square of $\tilde u^*$ as
$$
\tilde u_{rms}^*(z^+)=\sqrt{\overline{\tilde u^{*2}(T^+,z^+)}}
$$
and assuming that
\begin{equation}\label{eqn:uStarMHM}
u^*_{\mathrm{MHM}}=\tilde u_{rms}^*(z^+)\frac{\tilde u^*(T^+,Z^+)}{\tilde u_{rms}^*(Z^+)}
\end{equation}
Note that the right hand-side of (\ref{eqn:uStarMHM}) is frequency modulated via $T^+,$ while MHM did not mention frequency modulation. The same is true for scale modulation via $xu_{\tau_L}(t)/\nu$ and $yu_{\tau_L}(t)/\nu.$ This would make a difference where, for example, the spectra are concerned. However, for random mean square this difference does not matter.  
Then (\ref{eqn:FullUprimeModulationA}) takes the form
\begin{equation}\label{eqn:FullUprimeModulationB}
{u'}_p^+=u^*_{\mathrm{MHM}}\frac{\tilde u_{rms}^*(Z^+)}{\tilde u_{rms}^*(z^+)}\left(1+\frac{u_{\tau_L}'(t)}{\bar u_\tau}\right)+U^*\left(Z^+\right)\frac{u_{\tau_L}'(t)}{\bar u_\tau}+U^*\left(Z^+\right)-\frac{\bar u}{\bar u_\tau}
\end{equation}

The second difficulty is that if we now try to relate the modulation factor and the superposition term to ${u'}_{OL}^+$, at least one of them will have to depend on ${u'}_{OL}^+$ nonlinearly, while in (\ref{eqn:MathisEqTwoPointOne}) these relations are both linear. Noting that 
$$
Z^+=z^+\left(1+\frac{u_{\tau_L}'}{\bar u_\tau}\right)
$$
and linearizing under the assumption that $u_{\tau_L}'\ll\bar u_\tau$ gives  
\begin{equation}\label{eqn:LinearisedUprimeModulation}
{u'}_p^+ \approx u^*_{\mathrm{MHM}}\left[1+\left(1+\frac{z^+}{\tilde u_{rms}^*(z^+)}\frac{d\tilde u_{rms}^*(z^+)}{dz^+}\right)\frac{u_{\tau_L}'(t)}{\bar u_\tau}\right]+ 
\left(U^*(z^+)+z^+\frac{dU^*(z^+)}{dz^+}\right)
\frac{u_{\tau_L}'(t)}{\bar u_\tau}
\end{equation}

%
%

From the relationship between $\tau$ and $u_\tau$ after linearization one gets
$$
\frac{\tau_L'(t)}{\bar\tau}\approx2\frac{u_{\tau_L}'(t)}{\bar u_\tau}.
$$
From Section~\ref{sec:SkinFriction} we also have
$$
\frac{\tau_L'(t)}{\bar\tau}=\alpha {u'}_{OL}^+.
$$
Hence, 
\begin{equation}\label{eqn:UTauAlphaUOLRel}
\frac{u_{\tau_L}'(t)}{\bar u_\tau}\approx\frac12 \alpha {u'}_{OL}^+.
\end{equation}
With this substitution, comparing (\ref{eqn:MathisEqTwoPointOne}) and (\ref{eqn:LinearisedUprimeModulation}) gives
\begin{equation}\label{eqn:alphau}
\alpha_u(z^+)=\frac12\left(U^*(z^+)+z^+\frac{dU^*(z^+)}{dz^+}\right)\alpha
\end{equation}
and
\begin{equation}\label{eqn:betau}
\beta_u(z^+)=\frac12\left(1+\frac{z^+}{\tilde u_{rms}^*(z^+)}\frac{d\tilde u_{rms}^*(z^+)}{dz^+}\right)\alpha.
\end{equation}

\begin{figure}
  \begin{center}
    \raisebox{2.5in}{$(a)$} 
    \psfrag{xlab}{$z^+$}
    \psfrag{ylab}[Bc][][1][-90]{$\alpha_u$}
    \psfrag{leg1}[Br]{Theory}
    \psfrag{leg2}[Br]{Experiment}
    \includegraphics[width=0.75\textwidth]{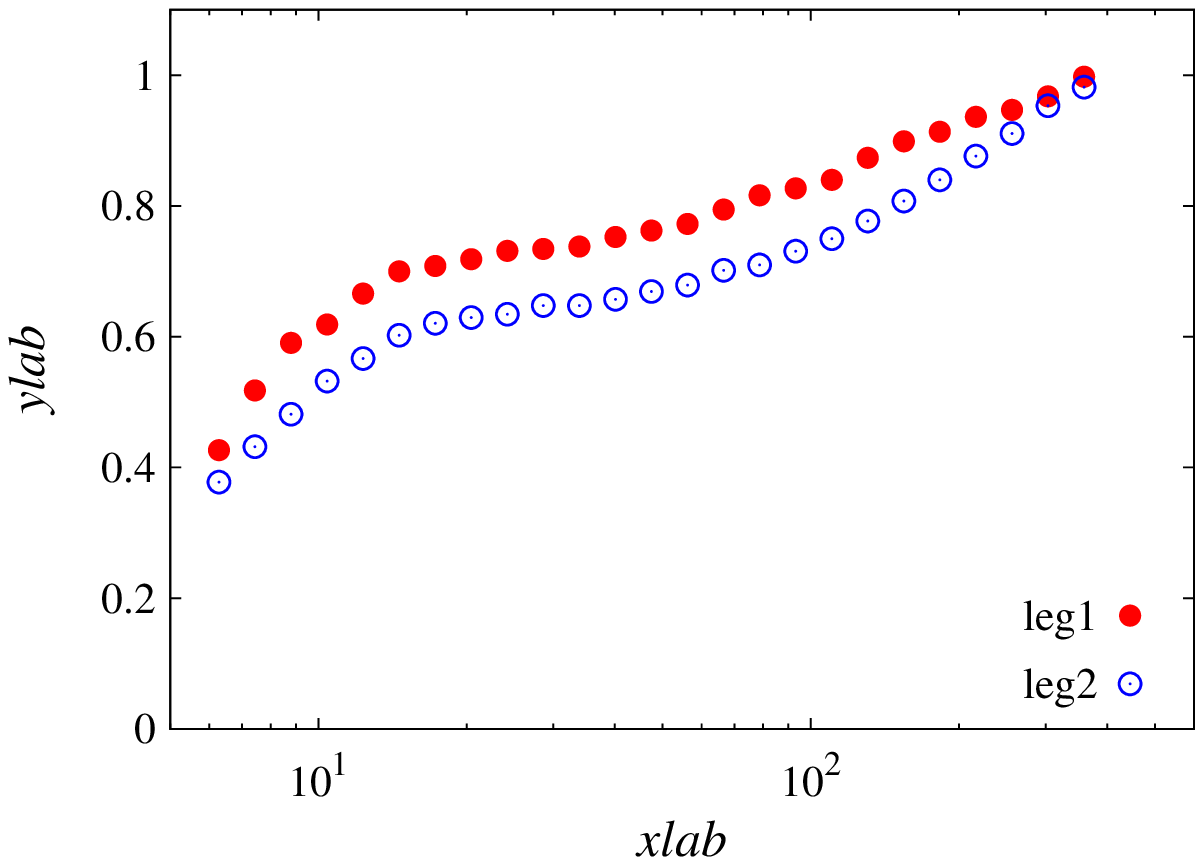}

    \raisebox{2.5in}{$(b)$} 
    \psfrag{ylab}[Bc][][1][-90]{$\beta_u$}
    \psfrag{leg1}[Bl]{Theory}
    \psfrag{leg2}[Bl]{Experiment}
     \includegraphics[width=0.75\textwidth]{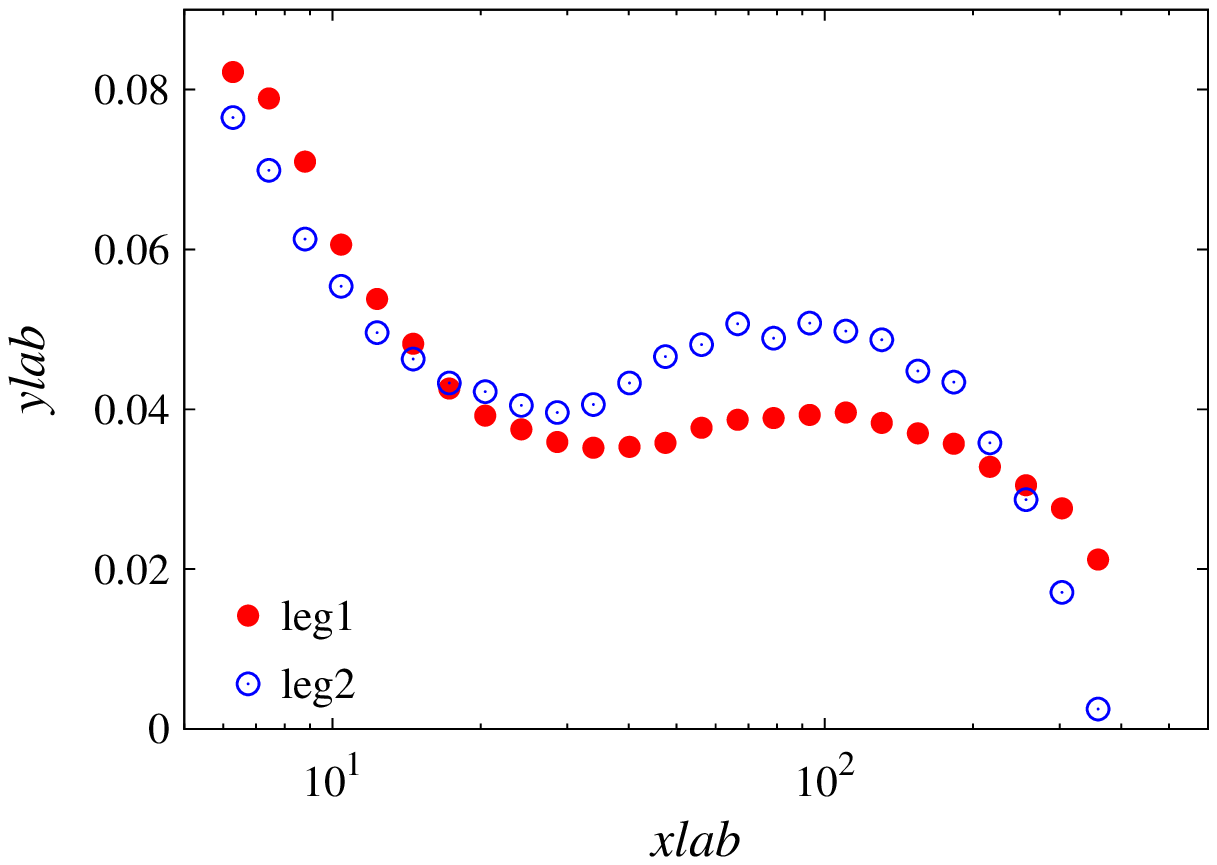}
  \end{center}
  \caption{Comparison of the theoretical expression of,
    $(a)$  $\alpha_u$, 
    and $(b)$ $\beta_u$, respectively from
    equations~(\ref{eqn:alphau}) and~(\ref{eqn:betau}) with values found
    experimentally in Mathis \etal~(2011).} 
  \label{fig:alpha}
\end{figure}

The universal mean profile $U^*(z^+)$ and the random mean square $\tilde u_{rms}^*(z^+)$ of the universal fluctuation $\tilde u^*(\dots,z^+)$ should, in principle, be adjusted so as to fit experimental results better over a range of Reynolds numbers. The most appropriate way would be of course to find the universal velocity $u^*$ by fitting (\ref{UniversalTotalVelocity}) directly to experimental data, without introducing fluctuations and without linearization. However, due to the linearisation the results are approximate in any case. For the first check of the theory we took $U^*$ and $\tilde u_{rms}^*$ to be equal to the mean profile in the calibration experiment and the random mean square of the universal signal of Mathis \etal~(2011) respectively, and we used $\alpha=0.0898$ from the same source.     
Comparison of (\ref{eqn:alphau}) and (\ref{eqn:betau}) with the
experimental results\footnote{Note that the definition of $\alpha$ ($\alpha_u$) in MHM should be corrected to include
  the ratio of random mean squares and should be $\alpha=\mathop{max}
  R(u_L^+,{u'}_{OL}^+)(u_L^+)_{r.m.s}/({u'}_{OL}^+)_{r.m.s}$. The
  plots were built with this correction.} 
 described in Mathis \etal~(2011) is shown in  Figures~\ref{fig:alpha}$(a)$ and~\ref{fig:alpha}$(b)$.

\section{Skin friction superposition coefficient}

So far, the skin friction superposition coefficient $\alpha$ was considered as a constant measured in experiment independently of all other quantities. Given its value, our theory established the relationship between the universal mean velocity profile and universal random-mean-square velocity fluctuation profile and the functions $\alpha_u(z^+)$ and $\beta_u(z^+).$ For the purposes of comparisons in Figures~\ref{fig:alpha}$(a)$ and~\ref{fig:alpha}$(b)$ the experimentally measured value of $\alpha$ was used. 

One further step can be made if, in the spirit of the classical method
of matched asymptotic expansions, one recognises that the logarithmic
layer is, in fact, a part of the inner distinguished limit\footnote{It
  is also the part of the outer distinguished limit and, as such, is the region of overlap of the two distinguished limits.}. Because of that, the relationship (\ref{eqn:LinearisedUprimeModulation}), which is valid in the inner limit should also be valid in the logarithmic layer. 
  In MHM  the large scale signal ${u'}^+_{OL}$ was chosen to be the low-pass filtered velocity ${u'}_p^+$ at a position in the logarithmic layer at $z^+=z^+_{OL}=3.9\Rey_\tau^{1/2}$. Applying the low-pass filter to (\ref{eqn:LinearisedUprimeModulation}) removes the fast-varying term. Substituting $z^+=z^+_{OL}$ and using (\ref{eqn:UTauAlphaUOLRel}) gives,    

$$
{u'}^+_{OL} \approx  
\left(U^*(z^+)+z^+\frac{dU^*(z^+)}{dz^+}\right)_{z^+=z^+_{OL}}
\frac12\alpha {u'}^+_{OL}
$$
Cancelling out ${u'}^+_{OL}$ one obtains

$$
\alpha=2/\left(U^*(z^+)+z^+\frac{dU^*(z^+)}{dz^+}\right)_{z^+=z^+_{OL}} 
$$

Since $z^+_{OL}$ is inside the logarithmic layer, one can take $U^*(z^+)=\frac1{\kappa}\ln{z^+}+B,$ with $\kappa=0.384$ and $B=4.17$ (\citealt{MonkewitzPF2007}).
This gives
\begin{equation}\label{eqn:alpha}
\alpha=2/\left(\frac1{\kappa}\ln{z^+_{OL}}+B+\frac1{\kappa}\right).
\end{equation}

\begin{figure}
  \begin{center}
    \psfrag{xlab}{$z^+_{OL}$}
    \psfrag{ylab}[Bc][][1][-90]{$\alpha_u$}
    \psfrag{leg1}[Br]{$\alpha =
      2/\left(\displaystyle\frac1{\kappa}\ln{z^+_{OL}}+B+\frac1{\kappa}\right)$}   
    \psfrag{leg2}[Br]{Experiment}
    \psfrag{leg3}[Br]{DNS}
    \includegraphics[width=0.75\textwidth]{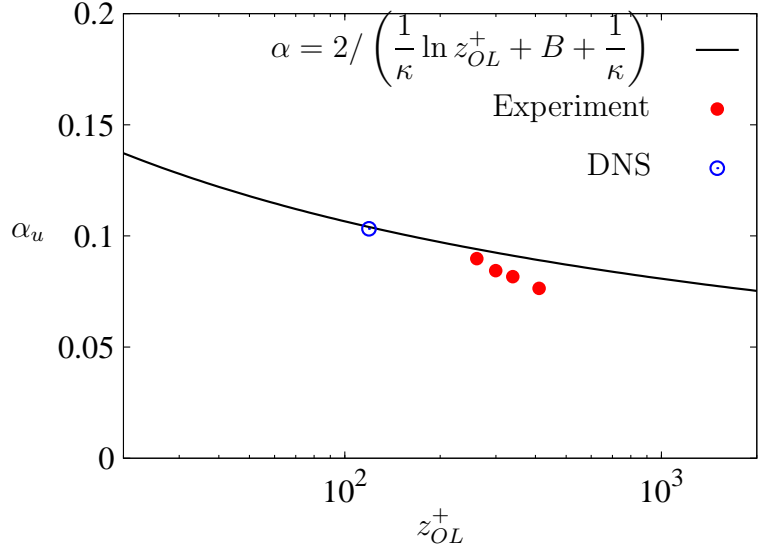}
  \end{center}
  \caption{Dependence of $\alpha$ with the wall-normal
    location of $z^+_{OL}$.} 
  \label{fig:alphatrend}
\end{figure}




Figure~\ref{fig:alphatrend} shows the comparison for $\alpha$
calculated from  (\ref{eqn:alpha}) along with values obtained from
experiments and DNS data. 
It is seen that the value of $\alpha$ derived here depends on the
choice of the wall-normal location $z^+_{OL}$ at which the
large-scales signal is taken. This result is consistent with MHM since the coefficient $\alpha$ is 
proportional to the coefficient correlation $\text{max}\{R(u_\tau,
u_{OL})\}$, and is thus known to depend on the location of the
outer-probe (Brown \& Thomas; Marusic \& Heuer 2007).

\nocite{marusic07_PRL,brown77_PoF}


\section{Concluding remarks}

In this paper we have investigated the simplest theoretical
description of the phenomenon of modulation of near-wall turbulence by
large scale structures associated with the logarithmic layer. In this description the effect of large scale structures is limited to a quasi-steady response of the near-wall turbulence to slow large-scale fluctuations of the skin friction. The most natural and compact form of expressing this mechanism is given by (\ref{eqn:TauModulated}) and (\ref{UniversalTotalVelocity}), which are simply the usual Reynolds-number-independent representation of total  skin friction and velocity in wall variables but with mean skin friction replaced by large-scale low-pass-filtered fluctuating skin friction. This mechanism manifests itself both in amplitude modulation and superposition, introduced in previous works, and frequency and wavenumber modulation which has not been considered before.
 When rewritten in terms of fluctuations and linearised with an assumption that the large-scale fluctuation of the skin friction is small as compared to the mean skin friction, our formulae coincide with the formulae proposed by MHM. However,  an additional feature here is that in the proposed theory the skin friction superposition and modulation constants $\alpha$ and $\beta$ and the analogous functions $\alpha_u(z^+)$ and $\beta_u(z^+)$ for the velocity distributions are expressed explicitly via the universal mean velocity profile, and the universal random mean square fluctuation velocity profile. 
In MHM these were obtained purely from experiments.
The predicted values of $\alpha$ and $\beta$ agree reasonably well with the values obtained experimentally. 

The agreement for $\alpha_u(z^+)$ and $\beta_u(z^+)$ is also encouraging, especially if one takes into account the error introduced by the linearization. The linearization was needed, of course, only in order to be able to make the comparisons. It is more appropriate and might be easier to use the exact theoretical formulae. As far as the comparisons are concerned, one should also note that the theory is applicable only in the inner layer, including the log layer, and that the theoretical results expressed in wall units are independent of the Reynolds number. The way $\beta_u$ was defined in experiments (MHM) was such that $\beta_u(z^+=3.9\Rey_\tau^{1/2})=0,$ that is $\beta_u$ is dependent on the Reynolds number at least near the largest values of its argument, which, therefore, is outside the region of applicability of the theory. Overall, the comparisons strongly support the theory.

It is of interest to note the simple physical mechanism that leads to the reduction, and even possible change of sign, of the amplitude modulation of the turbulence kinetic energy of small-scale fluctuations as the distance from the wall increases. An increase in the large-scale velocity has two effects. First, it increases the skin friction, and that leads to an increase in the turbulence kinetic energy of small-scale fluctuations. Hence, the first effect creates a positive correlation between the large-scale velocity and small-scale turbulence intensity. Second, an increase in the skin friction leads to a decrease in the thickness of the inner region. As a result, at a fixed physical distance from the wall the value of $z^+$ increases. If this point happens to be in the region where the turbulence kinetic energy of small-scale fluctuations decreases with $z^+$, this second effect will decrease the degree of amplitude modulation, thus contributing negatively to the correlation. More accurate information on the universal distribution of the turbulence kinetic energy of small-scale fluctuations would be needed in order to confirm if this mechanism can indeed lead to negative modulation at some distances from the wall.

The modulation theory makes it possible to extrapolate the results of observations at moderate Reynolds numbers to much higher values of Reynolds number. For example, it is well known that riblets reduce turbulent skin friction. The drag-reducing effect is strongest when riblet dimensions expressed in wall units attain an optimal value. At high Reynolds number, however, the large-scale structures will result in slow variations of the skin friction, and, hence, this implies that the effective riblet dimension in wall units cannot remain optimal at all times, thus leading to the decrease of the effectiveness of riblets at higher Reynolds numbers. By predicting the magnitude of the large-scale structures, say, using the experimental fit of \cite{Alfredsson2011ETC13}, and then using the known dependence of the drag reduction effect on riblet geometry for moderate Reynolds number (\citealt{GarciaMayoralJimenez2011}) and the formulae derived in the present study, it is possible to estimate quantitatively the change in the drag reduction due to an increase in  Reynolds number. The same idea applies to drag reduction by wall oscillations. Of course, this implies that the modulation effect, confirmed by comparisons in the present work only for the flow past a flat solid wall, will also remain the only effect of Reynolds number in flows with drag reduction. These considerations show, however, the importance of further studies of the modulation effect.

\vspace{1cm}

The authors gratefully acknowledge the financial support from EPSRC through grant EP/G060215/1, together with Airbus Operations Ltd and EADS UK Ltd, and the Australian Research Council.

\bibliographystyle{jfm}


\begin{thebibliography}{24}

\bibitem[Abe, Kawamura,  \& Choi (2004)]{AbeKawamuraChoiASMEFluidEng2004}
  \textsc{Abe, H., Kawamura, H. \& Choi, H.} 2004 Very large-scale structures and their effects on the wall
  shear-stress fluctuations in a turbulent channel flow up to $\Rey_\tau = 640.$ \emph{Trans. ASME: J. Fluid
    Engng} \textbf{126}, 835--843.

\bibitem[Adrian, Meinhart, \& Tomkins, (2000)]{AdrianMeinhartTomkins2000}
  \textsc{Adrian, R. J., Meinhart, C. D. \& Tomkins, C. D.} 2000 Vortex organization in the outer region of
  the turbulent boundary layer. \emph{J.~Fluid Mech.} 422, 1--54.

\bibitem[Alfredsson, \" Orl\" u,  \& Segalini (2011)]{Alfredsson2011ETC13}
  \textsc{Alfredsson, P. H., \" Orl\" u, R.,  \& Segalini, A. S.} 2011 A new formulation for the streamwise turbulence intensity distribution. 
{\em Proc. 13th European Turbulence Conference (ETC13)}, Warsaw, Poland. 
  
  \bibitem[Brown \& Thomas(1977)]{brown77_PoF}
{\sc Brown, G.~L. \& Thomas, A.~S.~W.} 1977 Large structure in a turbulent
  boundary-layer. {\em Phys. Fluids\/} {\bf 20}~(10), S243--S251.

\bibitem[del \' Alamo \& Jim\' enez (2003)]{delAlamoJimenezPF2003}
  \textsc{del \' Alamo \& J. C., Jim\' enez} 2003 Spectra of the very large anisotropic scales in turbulent
  channels. \emph{Phys.~Fluids} \textbf{15 (6)}, L41--L44.

\bibitem[del \' Alamo et al (2004)]{delAlamoJimenezEtAlJFM2004}
  \textsc{del \' Alamo, J. C., Jim\' enez, J., Zandonade, P. \& Moser, R. D.} 2004 Scaling of the energy spectra
  of turbulent channels. \emph{J.~Fluid Mech.} \textbf{500}, 135--144.

\bibitem[Ganapathisubramani \emph{et al.} (2005)]{GanapathisubramaniEtAlJFM2005}
  \textsc{Ganapathisubramani,  B.,  Hambleton,  N.  Hutchins  W.  T.,  Longmire,  E.  K.  \&  Marusic,  I.}
  2005 Investigation of large-scale coherence in a turbulent boundary layer using two-point
  correlation. \emph{J.~Fluid Mech.} \textbf{524}, 57--80.

\bibitem[Ganapathisubramani, Longmire, \& Marusic (2003)]{BharathJFM2003}
  \textsc{Ganapathisubramani, B., Longmire, E. K. \& Marusic, I.} 2003 Characteristics of vortex packets
  in turbulent boundary layers. \emph{J.~Fluid Mech.} \textbf{478}, 35--46.


\bibitem[Garc\'ia-Mayoral \& Jim\'enez (2011)]{GarciaMayoralJimenez2011}
  \textsc{Garc\'ia-Mayoral, R. \& and Jim\'enez, J.} 2011 
  Drag reduction by riblets. \emph{Phil. Trans. R. Soc. A} 
  \textbf{369}, 1412--1427. 


\bibitem[Hambleton,  Hutchins, \&  Marusic  (2006)]{HambletonHutchinsMarusicJFM2006}
  \textsc{Hambleton,  W.  T.,  Hutchins,  N.  \&  Marusic,  I.}  2006  Simultaneous  orthogonal-plane  particle
  image velocimetry measurements in a turbulent boundary layer. \emph{J.~Fluid Mech.} \textbf{560}, 53--64.

\bibitem[Hoyas \&  Jim\' enez (2006)]{HoyasJimenezPF2006}
  \textsc{Hoyas,  S.  \&  Jim\' enez,  J.}  2006  Scaling  of  the  velocity  fluctuations  in  turbulent  channels  up  to $\Rey_\tau = 2003.$ \emph{Phys.~Fluids} \textbf{18}, 011702.

\bibitem[Hutchins \& Marusic (2007{\natexlab{{\em a\/}}})]{HutchinsMarusic2007a}
  \textsc{Hutchins, N. \& Marusic, I.} 2007 Evidence of very long meandering features in the logarithmic
  region of turbulent boundary layers. \emph{J.~Fluid Mech.} \textbf{579}, 1--28.

\bibitem[Hutchins \& Marusic(2007{\natexlab{{\em b\/}}})]{HutchinsMarusic2007b}
{\sc Hutchins, N. \& Marusic, I.} 2007{\natexlab{{\em b\/}}} Large-scale
  influences in near-wall turbulence. {\em Phil. Trans. R. Soc. Lond. A\/} {\bf
  365}, 647--664.

\bibitem[Marusic \& Heuer(2007)]{marusic07_PRL}
{\sc Marusic, I. \& Heuer, W.~D.~C.} 2007 {R}eynolds number invariance of the
  structure angle in wall turbulence. {\em Phys. Rev. Lett.\/} {\bf 99},
  114501.
  
\bibitem[Marusic, Mathis, \& Hutchins (2010)]{marusic10_SCI}
{\sc Marusic, I., Mathis, R. \& Hutchins, N.} 2010
  Predictive model for wall-bounded turbulent flow. {\em Science\/} {\bf
  329}~(5988), 193--196.

\bibitem[Marusic, Mathis, \& Hutchins (2011)]{MarusicETC} 
  \textsc{Marusic, I., Mathis, R., \& Hutchins, N.} 2011 
  A wall-shear stress predictive model.  {\em Proc. 13th European Turbulence Conference (ETC13)}, Warsaw, Poland.

\bibitem[Mathis, Hutchins,  \& Marusic (2009)]{MathisHutchinsMarusicJFM2009}
  \textsc{Mathis, R., Hutchins, N. \& Marusic, I.} 2009 Large-scale
  amplitude modulation of the small-scale structures in turbulent
  boundary layers. \emph{J.~Fluid Mech.} \textbf{628}, 311--337. 

\bibitem[Mathis, Hutchins, \& Marusic (2011)]{MathisJFM2011} 
  \textsc{Mathis, R., Hutchins, N., \& Marusic, I.} 2011 A predictive
  inner-outer model for streamwise turbulence statistics in
  wall-bounded flows. \emph{J.~Fluid Mech.} \textbf{681}, 537--566. 

\bibitem[McKeon \& Morrison (2007)]{McKeonMorrisonPTRS2007}
  \textsc{McKeon, B. J. \& Morrison, J. F.} 2007 
  Asymptotic scaling in turbulent pipe flow. \emph{Phil. Trans. R. Soc. A} 
  \textbf{365}, 771--788.
 

\bibitem[Monkewitz, Chauhan, \& Nagib (2007)]{MonkewitzPF2007}
{\sc Monkewitz, P. A., Chauhan, K. A., \& Nagib, H. M.} 2007. Self-consistent high-Reynolds-number asymptotics for
  zero-pressure-gradient turbulent boundary
  layers. {\em Phys.~Fluids} {\bf 19}, 115101. 

\bibitem[Morrison (2007)]{MorrisonPTRS2007} 
  \textsc{Morrison, J. F.} 2007 
  The interaction between inner and outer regions of turbulent wall-bounded flow. 
  \emph{Phil.~Trans.R.~Soc.~A} 
  \textbf{365}, 683--698. 
  
  \bibitem[\"Orl\"u \& Schlatter (2011)]{OrluSchlatter2011}
    \textsc{\"Orl\"u, R., \& Schlatter, P.} 2011. 
On the fluctuating wall-shear stress in zero-pressure-gradient turbulent boundary layers. 
\emph{Phys.~Fluids} \textbf{23}, 021704. 



\end{thebibliography}

\end{document}